\documentclass[a4paper,fleqn,usenatbib]{mnras}

\usepackage[T1]{fontenc}
\usepackage{ae,aecompl}

\usepackage{graphicx}	
\usepackage{amsmath}	
\usepackage{amssymb}	
\usepackage{multicol}        
\usepackage{bm}		
\usepackage{color}
\usepackage{xcolor}

\title[Spots and flares in hot main sequence stars]
{Spots and flares in hot main sequence stars}

\author[L. A. Balona]{\
L. A. Balona \thanks{E-mail: lab@saao.ac.za} 
\\
South African Astronomical Observatory, 
P.O. Box 9, Observatory 7935, South Africa}

\begin{document}

\date{Accepted .... Received ...}

\pagerange{\pageref{firstpage}--\pageref{lastpage}} \pubyear{2011}

\maketitle

\label{firstpage}

\begin{abstract}
About 22000 {\it Kepler} stars and nearly 60000 {\it TESS} stars from sectors 
1--24 have been classified according to variability type.  A large proportion 
of stars of all spectral types appear to have periods consistent with the
expected rotation periods.  A previous analysis of A and late B stars
strongly suggests that these stars are indeed rotational variables.  In this
paper we have accumulated sufficient data to show that rotational modulation
is present even among the early B stars.  A search for flares in {\it TESS}
A and B stars resulted in the detection of 110 flares in 68 stars.  The flare 
energies exceed those of typical K and M dwarfs by at least two orders of 
magnitude.  These results, together with severe difficulties of current models 
to explain stellar pulsations in A and B stars, suggest a need for revision of 
our current understanding of the outer layers of stars with radiative 
envelopes.  
\end{abstract}


\section{Introduction}

High-precision space photometry of upper main sequence stars show periodic or 
quasi-periodic variations with periods consistent with the expected rotational 
periods of these stars \citep{Balona2013c, Balona2016a, Balona2017a, 
Balona2019c}.  High-resolution spectroscopic time series of Vega (A0V) indicates
the presence of a spotted stellar surface\citep{Bohm2015}, providing
independent confirmation of the photometric results.  

In addition, early results from the {\it Kepler} mission \citep{Borucki2010} 
indicated the presence of flares associated with some A and late-B stars 
\citep{Balona2012c}. Further studies \citep{Balona2013c, Balona2015a, 
Balona2016e} seem to indicate that around 2.5\,percent of A stars flare with 
energies in the range $10^{35}$--$10^{36}$\,erg.  \citet{Pedersen2017} have 
argued that the flares are likely a result of cool flare stars in the same 
aperture or binary companions.  Direct evidence of possible X-ray flares in A 
stars have been reported by \citet{Schmitt1994}, \citet{Robrade2010}, while a 
flare on a B star has also been reported \citep{Yanagida2004, Yanagida2007}.

Rotational modulation and flares in A and B stars implies the presence of 
surface magnetic fields, contrary to the long-held view that it is not
possible for stars with radiative envelopes to host magnetic fields.
The Ap and Bp stars have strong global magnetic fields, but these are 
explained as being of fossil origin \citep{Mestel1967}.  Photometric 
studies referenced above indicate that rotational modulation is present in as 
many as 40\,percent of stars on the upper main sequence, most of which are not 
known Ap/Bp stars.

First results from the {\it Kepler} space mission on pulsations in
main-sequence A stars \citep{Grigahcene2010} already indicated a serious 
problem.  It turns out that nearly all $\delta$~Scuti stars have multiple low 
frequency pulsations which cannot be explained by current models 
\citep{Balona2014a, Balona2018c}.  A further surprise was the confirmation that 
many late-B stars pulsate with high frequencies (Maia variables, 
\citealt{Balona2015c,Balona2016c, Balona2020a}).  These are difficult to 
explain in terms of incorrect opacities alone \citep{Daszynska-Daszkiewicz2017b}.
Perhaps of even more significance is the fact that less than half of the
stars in the $\delta$~Sct instability strip pulsate.  Also it seems that the
$\gamma$~Dor variables may be just a subset of the $\delta$~Sct stars
\citep{Balona2018c}.  None of these findings are explained by current pulsation 
models. 

New ideas regarding the outer layers of stars in radiative envelopes have
recently emerged.  It has been suggested, for example, that generation of
magnetic fields by dynamo action may occur in sub-surface convective zones in 
A and B stars \citep{Cantiello2009, Cantiello2011,Cantiello2019}.  At the 
surface they give rise to bright starspots.  Also, it has been suggested
that differential rotation may act to provide dynamo-generated magnetic
fields in radiative zones \citep{Spruit1999, Spruit2002, Maeder2004}.

In this paper we report on further evidence for rotational modulation among
{\it TESS} A and B stars, indicating that starspots are common among all B
stars, including the early-type B stars.  We also report on a survey for 
flares in {\it TESS} stars on the upper main sequence. We argue that current 
ideas regarding the outer layers of stars in radiative equilibrium need to be revised.

\section{Data and methodology}

The data used here are the full four-year light curves from {\it Kepler} and
sectors 1--24 of {\it TESS} data.  In both cases the light curves are obtained 
using pre-search data conditioning (PDC) which corrects for time-correlated 
instrumental signatures in the light curves \citep{Jenkins2016}. All stars
with effective temperatures $T_{\rm eff} > 6000$\,K brighter than magnitude
12.5 were selected for the analysis of rotational modulation.  This results in 
5643 stars from {\it Kepler} and over 50000 stars from {\it TESS}.  In the 
search for flares, the uncorrected light curves were used and only stars
with $T_{\rm eff} > 7500$\,K were selected.

Visual inspection of the light curves and the Lomb-Scargle periodograms 
\citep{Scargle1982} of {\it Kepler} and {\it TESS} stars were used to assign 
variability types whenever appropriate.  The variability classification follows
that of the {\it General Catalogue of Variable Stars} (GCVS, 
\citealt{Samus2017}).  The only recognized class of rotational variable
among the A and B stars are the chemically peculiar $\alpha$~CVn and SX~Ari
classes.  A new ROT class has been added to describe any star in which the
variability is suspected to be due to rotation and not known to be Ap or Bp.  
Aided by suitable software, visual classification of over 100 stars an hour is 
possible.  In this way, several thousand stars with $T_{\rm eff} > 6000$\,K 
have been assigned the ROT type.

\section{Stellar parameters}

The most commonly used test for rotational modulation is comparison of the
rotation rate derived from the photometric frequency, $\nu_{\rm ROT}$, with 
that derived from the projected rotational velocity, $v\sin i$.  To derive
the equatorial rotational velocity, $v$, from $\nu_{\rm ROT}$ requires an
estimate of the stellar radius, $R$.  This can be done if we know the
effective temperature, $T_{\rm eff}$ and luminosity, $L/L_\odot$.  

The most precise method of deriving $T_{\rm eff}$ is by modelling absorption 
line profiles from medium- or high-resolution spectroscopy.  For A and B stars,
this involves fitting the H$\beta$ and/or H$\alpha$ line profiles using a 
suitable model atmosphere.  The resulting standard deviation in $T_{\rm eff}$ 
ranges from about 100\,K for A stars to about 1000\,K for early B stars.  
Spectroscopic estimates of $T_{\rm eff}$ exist for about 25\,percent of the 
sample considered here.  

The next best method is the use of narrow-band photometry.  This involves 
measuring the strength  of the H$\beta$ line (the Str\"{o}mgren $\beta$ index)
usually in conjunction with $uvby$ narrow-band photometry.  The value of
$T_{\rm eff}$ is obtained either by direct comparison with synthetic
photometry derived from model atmospheres or by using stars with known
$T_{\rm eff}$ \citep{Moon1985b, Gray1991, Napiwotzki1993, Smalley1993a, 
Balona1994a}.  Estimates of $T_{\rm eff}$ from Sloan $ugriz$
\citep{Brown2011a} are of this type and are available for most of the 
{\it Kepler} and {\it TESS} stars.  However, they cannot be used for stars
earlier than A0 because they lack $u$-band measurements.  Without the
$u$ band, it is impossible to distinguish between A and B stars of the same
colour.  Estimates of $T_{\rm eff}$ using narrow- and intermediate-band 
photometry are available for about 55\,percent of stars.

If neither spectroscopy or narrow-band photometry is available, wide-band 
photometry can be used to estimate $T_{\rm eff}$ provided that the reddening 
is known.  This method is used in 7\,percent of our sample.  Finally, if nothing 
else is available, a crude estimate of $T_{\rm eff}$ can be derived from the 
spectral classification together with suitable calibration such as the
calibration of \citet{Pecaut2013}.  This method was used for 18\,percent of 
the stars.

The stellar luminosity is best estimated from {\it Gaia} DR2 parallaxes  
\citep{Gaia2016, Gaia2018} in conjunction with reddening estimated using a 
three-dimensional map by \citet{Gontcharov2017} and the bolometric correction 
calibration by \citet{Pecaut2013}. From the error in the {\it Gaia} DR2 
parallax, the typical standard deviation in $\log(L/L_\odot)$ is estimated to 
be about 0.05\,dex, allowing for standard deviations of 0.01\,mag in the 
apparent magnitude, 0.10\,mag in visual extinction and 0.02\,mag in the 
bolometric correction in addition to the parallax error.

\begin{table}
\begin{center}
\caption{Number of ROT stars within the given $T_{\rm eff}$ range, $N_{\rm
ROT}$.  Also shown is the fraction of ROT stars, $f_{\rm ROT}$ within the 
range. $N_{\rm vsini}$ is the number of stars used to construct the $v\sin
i$ vs $v$ diagrams (Fig.\,\ref{distr} left panel) and $N_{\rm dist}$ is 
the number of stars used in obtaining the $v\sin i$ distribution 
(Fig.\,\ref{distr} right panel).}

\vspace{2mm}

\label{tab}
\begingroup
\setlength{\tabcolsep}{10pt} 
\renewcommand{\arraystretch}{1.3} 
\begin{tabular}{crrrr}
\hline
\multicolumn{1}{c}{$T_{\rm eff}$}   & 
\multicolumn{1}{c}{$N_{\rm ROT}$}   &
\multicolumn{1}{c}{$f_{\rm ROT}$}   &
\multicolumn{1}{c}{$N_{\rm vsini}$} &
\multicolumn{1}{c}{$N_{\rm dist}$} \\
\hline
   6000--7000  &   21835 & 0.50 & 3329 & 7128  \\
   7000--8000  &    3298 & 0.34 &  239 & 1459  \\
   8000--10000 &    2418 & 0.31 &  420 & 2068  \\
  10000--12000 &     529 & 0.40 &  205 &  866  \\
  12000--18000 &     341 & 0.37 &  219 & 1417  \\
  18000--30000 &     138 & 0.29 &   85 & 1481  \\
\hline                        
\end{tabular}
\endgroup
\end{center}
\end{table}

Table\,\ref{tab} lists the number of stars classified as ROT variables in
the given range of effective temperature.   Also shown is the percentage of
main sequence stars for which the ROT classification was assigned.  Note
that Be stars were excluded from the sample.  While the light variations in
Be stars can be interpreted as rotational modulation, the light amplitude is 
typically an order of magnitude larger than for non-Be stars 
\citep{Balona2020b}.  It is suggested that the cause of the variability are
co-rotating clouds which obscure a larger fraction of the photosphere than
starspots.  Because of the large amplitude, Be stars are disproportionately
represented among the ROT stars.  Since Be stars are rapid rotators, their 
inclusion leads to an over-estimate of the proportion of ROT stars with rapid 
rotation. Most Be stars are of early B type and this leads to a severe 
distortion of the velocity distribution for stars with 
$T_{\rm eff} > 18000$\,K.

\section{Results}

The photometric period, obtained from {\it Kepler}, {\it K2} and {\it TESS}
light curves together with the stellar radius is used to estimate the 
equatorial rotational velocity, $v$. If the variability is rotational
modulation, there should be a relationship between $v$ and the projected 
rotational velocity, $v\sin\,i$.  Since most stars will be 
observed roughly equator-on, one expects that most data points in the 
$v$--$v\sin i$ diagram will lie on or just below the straight line defining 
$\sin i = 1$.  There will be a diminishing scatter of points below the line due
to stars with lower angles of inclination, $i$.  Due to unavoidable errors,
some points are to be expected above the $v\sin i = 1$ line.

\begin{figure*}
\centering
\includegraphics[]{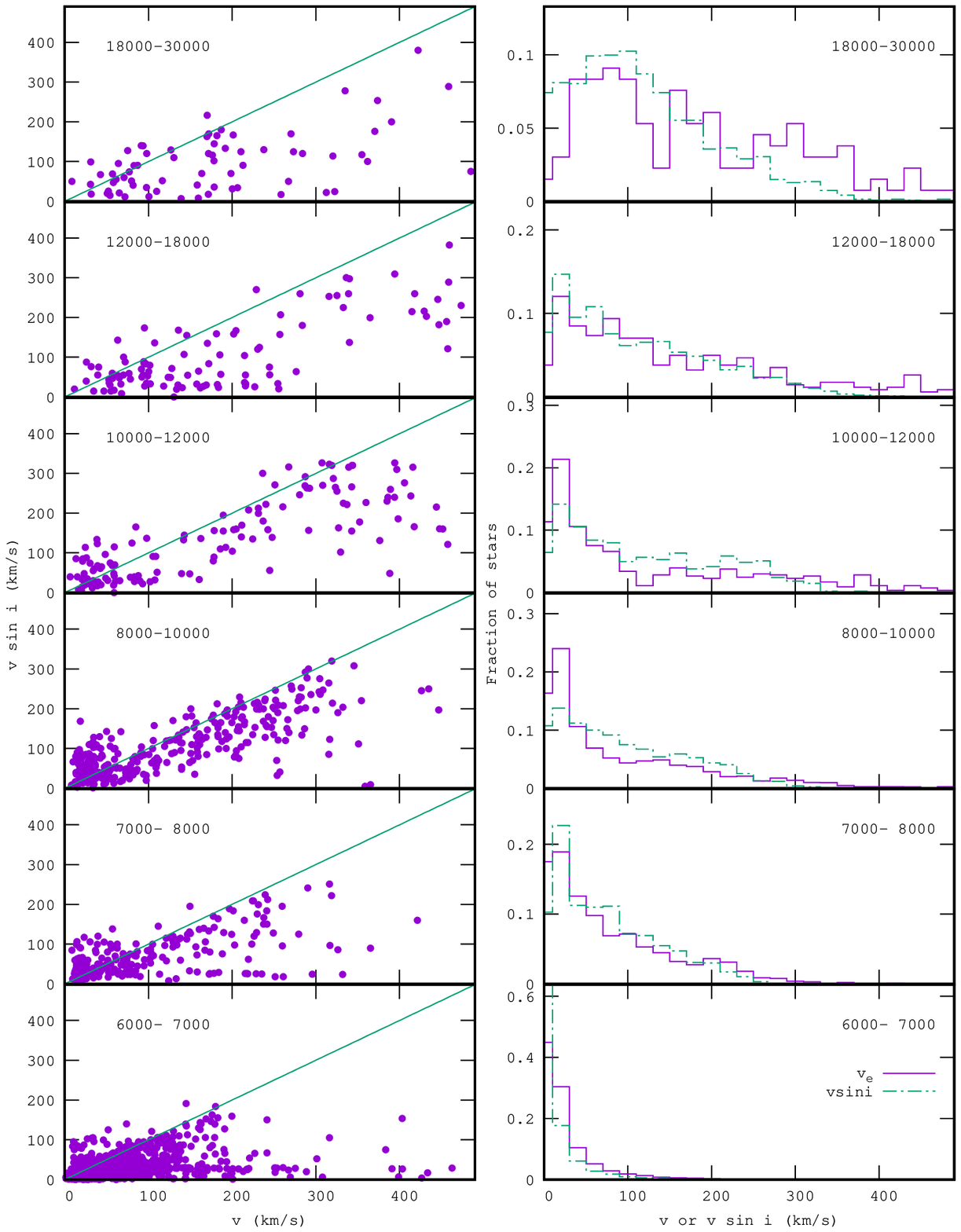}
\caption{Left panel: the relationship between the projected rotational 
velocity, $v\sin i$ and the equatorial rotational velocity, $v$ (estimated 
from the photometric frequency), for different ranges of $T_{\rm eff}$
(labeled).  The straight line corresponds to $v = v\sin i$.  Right panel: the 
distribution of equatorial rotational velocity (solid violet) and $v \sin i$ 
(dashed blue) for different ranges of $T_{\rm eff}$.  In both cases Be stars 
are omitted.}
\label{distr}
\end{figure*}

Projected rotational velocities are not available for every star for
which $v$ has been estimated.  In the left panel of Fig.\,\ref{distr}, the 
$v$--$v\sin i$ diagram is shown for all stars for which measurements are 
available.  It is clear that the expected distribution of points is present 
from the F stars to the early B stars.  This justifies the original assumption 
that the periodic light variation is due to rotation.  

As expected, nearly all stars with $v > 60$\,km\,s$^{-1}$ lie on or below the 
$\sin i = 1$ line.  For low rotation rates, ever increasing observational
precision is required to determine whether or not $v\sin i < v$.  Often, 
$v\sin i$ values are truncated at some positive number corresponding to the 
resolution limit of the instrument.  An important factor is that it becomes
increasingly difficult to distinguish between binarity and rotation at low
frequencies.  For example, amplitude variability, which is a typical attribute 
of rotational modulation, is not so easily detected.  Thus one may expect
significant contamination from binaries at low rotation rates.  These
factors are probably responsible for the increased scatter in this region.

A more rigorous test can be made by considering the rotational velocity
distribution for main-sequence stars within a limited $T_{\rm eff}$ range.
The rotational velocity distribution is the relative number of stars at a
given rotational velocity.  This is an important quantity which provides 
information on the physics of stellar rotation.  The test involves the
comparison of the $v$ distribution with the $v\sin i$ distribution.   These
two distributions should be similar, though not identical due to variation
of the inclination of the rotation axis.  Close agreement is expected because 
most stars would be viewed equator-on.  This is a far more rigorous test 
because it involves not just comparison of $v$ and $v\sin i$ for the same 
star, but also tests whether the detailed distribution of $v$ corresponds
closely to that of $v\sin i$.

It might be thought that one could derive $v$ from $v\sin i$ by deconvolution 
assuming random orientation of the axis of rotation.  In this way one could 
compare the photometric and spectroscopic $v$ directly. The reason why this
has not been done is that the stars from which the photometric $v$ are derived 
cannot have a random axis of rotation.  Clearly, if a star is nearly pole-on, 
no rotational modulation will be detected.  This is, of course, not true for 
$v\sin i$ since rotational broadening can be made for any inclination angle.  
Therefore the distribution of $v$ from deconvolution of $v\sin i$ cannot match 
the photometric $v$ distribution.  For the same reason, testing the
distribution of $i$ is not possible.

Obtaining the distribution requires sufficient numbers of stars within the
chosen range of $v$ or $v\sin i$ in order to be statistically meaningful.
The number of stars for which photometric $v$ measurements are available
is sufficiently large for this purpose.  If the corresponding values of $v
\sin i$ are restricted to the same stars for which $v$ is available, the
numbers would be too small.  Fortunately, it is not 
necessary to impose this restriction because it is reasonable to assume that 
the $v\sin i$ distribution will be the same for any set of main-sequence stars 
within the chosen range of $T_{\rm eff}$.  In other words, one may select
any set of main sequence stars with known $v\sin i$ within the required 
effective temperature range.

In the right panel of Fig.\,\ref{distr}, the $v$ and $v\sin i$ distributions 
for stars in six temperature ranges are shown.  There is good agreement for 
all temperature ranges, reinforcing the results derived from the $v$--$v\sin i$
diagrams. 

\section{{\it TESS} early-type flare stars}

In our catalog of nearly 60000 stars classified for variability, there
are 14495 stars with $T_{\rm eff} > 7500$\,K. Flares are difficult to 
detect in eclipsing binaries and other types of variable with high
amplitude.  Excluding these stars results in 6072 A and 1616 B stars.  This
sample was searched for flares by visual inspection.

\begin{figure*}
\centering
\includegraphics[scale=0.7]{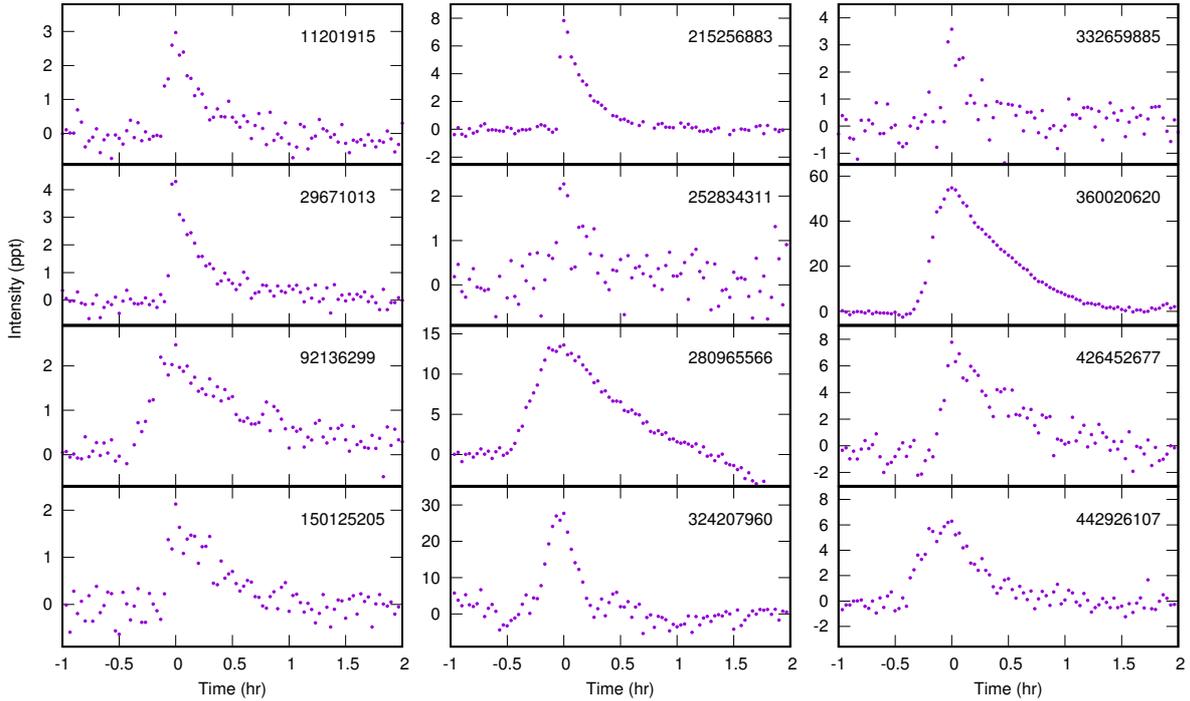}
\caption{Examples of A star flares in the {\it TESS} data.  The time origin
is the time of maximum flare intensity.  Time is in hours and intensity if
parts per thousand.}
\label{flarefig}
\end{figure*}

The 68 stars in Table\,\ref{flaretab} appear to have flare-like events,
examples of which are shown in Fig.\,\ref{flarefig}. A substantial proportion 
of the flare stars are X-ray sources.  Multiple flares are visible in 23 
stars, giving a total of 110 flare events.  Not included here are the {\it TESS} 
A-type flare stars TIC\,118327563 and TIC\,224244458 \citep{Balona2019a}. The 
former is a sdB star and the latter is an SX~Ari variable (Bp star).  Also 
excluded is the $\delta$~Sct flare star TIC\,439399707 \citep{Balona2019b} and 
the Be X-ray source TIC\,207176480 \citep{Balona2020b}.

The number of {\it TESS} A-type stars which appear to flare constitute about
1.1\,percent of the sample of A stars which were examined, which is less
than half of the 2.5\,percent flare incidence among {\it Kepler} A stars
reported by \citet{Balona2013c} and \citet{Balona2015a}.  This can be
understood given the fact that the long-cadence {\it Kepler} data span 4
years, while the {\it TESS} data mostly span a few months and always less
than one year.  There are 61 {\it Kepler} A stars known to flare
\citep{Balona2015a}.  The additional 68 flare stars reported here do not
include any of the {\it Kepler} stars, bringing the total of flaring A stars
to 129. 

Whereas the {\it Kepler} pixel size is 4\,arcsec, the {\it TESS} pixel size
is 21\,arcsec.  This means that the probability of a flare originating in
a star other than the A star is much larger for {\it TESS}.  The fields of
all 68 stars were examined and in each case the A star is by far the
dominant optical source.  Any cool star in the aperture would need to be of
comparable brightness to the A star for the flare to be detected.  A cool 
dwarf within the same aperture would be among the nearest stars and would have 
long ago been catalogued.  Therefore the source of the flare would need to be 
an exotic faint object, a companion of the A star or the A star itself.

The flare energy was estimated by integrating the light curve under the
flare. In cool dwarfs and the Sun, most of the flare energy is radiated in the 
UV.  The UV contribution in {\it Kepler} and {\it TESS} light curves is 
negligible, and therefore the estimated flare energy is likely to be an
underestimate.  As can be seen from Table\,\ref{flaretab}, the typical flare 
energy is $10^{35}$--$10^{36}$\,erg, which is about 10-1000 times more 
energetic than flares in cool dwarfs.  The most energetic flare in cool
dwarfs detected in a survey of {\it TESS} stars by \citet{Maximilian2020}
has an energy of $10^{34.7}$\,erg.  It is difficult to understand how a unique 
event like this can be repeated in over 1\,percent of A stars.  Multiple
flares in the same star, all with energies exceeding the highest ever seen,
also would seem to rule out a cool companion.

\begin{table*}
\begin{center}
\caption{Flare stars detected from visual inspection of {\it TESS} light
curves.  The TIC and HD numbers are followed by the assigned variability
class.  This is followed by the effective temperature (K) and luminosity.  The
time of peak flare intensity, $t_{\rm max}$, is relative to BJD\,2458000. 
The flare energy, $E$ (erg), is followed by the relative peak flare intensity,
$\tfrac{\Delta F}{F}$.  The number of flares, $N_{\rm Fl}$, is followed by a 
reference to X-ray detection and the spectral type.}
\label{flaretab}

\vspace{2mm}

\resizebox{0.77\textwidth}{!}{
\begingroup
\setlength{\tabcolsep}{10pt} 
\renewcommand{\arraystretch}{1.2} 
\begin{tabular}{rllrrrrrrrl}
\hline
\multicolumn{1}{c}{TIC}   & 
\multicolumn{1}{c}{HD}   & 
\multicolumn{1}{c}{Type}   & 
\multicolumn{1}{c}{$T_{\rm eff}$}   & 
\multicolumn{1}{c}{$\log L/L_\odot$}& 
\multicolumn{1}{c}{$t_{\rm max}$}   &
\multicolumn{1}{c}{$\log E$}   & 
\multicolumn{1}{c}{$\log\tfrac{\Delta F}{F}$}   & 
\multicolumn{1}{c}{$N_{\rm Fl}$}   & 
\multicolumn{1}{c}{Ref.}   & 
\multicolumn{1}{c}{Sp.Type}   \\
\hline
 11201915   & 37410      &        &   7966 &   1.26 &  486.73 &    35.3 & -2.528 & 1 & 1 &  kA4hA2VmA7   \\ 
 11895653   & 103287     &        &   9650 &   1.80 &  909.93 &    36.1 & -2.783 & 3 & 2 &  A0Ve+K2V SB  \\    
 22562087   & 107143     & ROT    &   8234 &   1.20 &  588.46 &    35.3 & -2.457 & 1 &   &  A1V          \\ 
 25424318   & 111608     & ROT    &   9155 &   1.40 &  573.68 &    35.9 & -1.972 & 1 &   &  A1IV         \\
 26893151   & 11060      & ROT    &   8494 &   1.18 &  780.81 &    36.3 & -2.196 & 2 & 3 &  A0           \\  
 28643592   & 174830     & ROT    &   7953 &   1.63 &  689.40 &    35.6 & -2.476 & 1 &   &  A2           \\
 29671013   & 200052     &        &   8892 &   1.47 &  336.79 &    35.7 & -2.367 & 1 &   &  A5V:pSiMg    \\
 30052567   & 76516      & ROT    &   8971 &   1.42 &  538.25 &    36.3 & -2.067 & 1 &   &  A0V          \\
 34404183   & 152384     & ROT    &   9096 &   1.55 &  648.00 &    36.4 & -2.552 & 1 &   &  A0V          \\  
 50624799   & 36118      & ROT    &  11183 &   1.80 &  489.77 &    36.7 & -2.012 & 1 & 4 &  B9V          \\  
 55219038   & 43620      & ROT    &   8476 &   1.43 &  862.38 &    35.5 & -2.534 & 1 &   &  A2           \\
 75873633   & 133574     & ROT    &   7078 &   0.92 &  618.67 &    35.6 & -2.128 & 2 &   &  A9/F0V       \\
 92136299   & 222661     & ROT    &  10618 &   1.73 &  380.64 &    36.2 & -2.607 & 3 & 5 &  B9.5IV       \\ 
 94336006   & 24300      &        &  13520 &   2.35 &  794.21 &    36.7 & -2.453 & 1 &   &  B8III?       \\
 125958765  & 154426     & ROT    &   7915 &   1.24 &  649.17 &    36.1 & -1.984 & 1 & 3 &  A7III        \\  
 142268253  & 16754      & ROT    &   8997 &   1.40 &  397.15 &    35.4 & -2.915 & 2 & 2 &  A1Va         \\   
 142457761  & 90759      &        &   8778 &   1.12 &  709.40 &    35.7 & -2.231 & 2 &   &  A2           \\
 147622676  & 94660      & ACV    &   9544 &   1.77 &  575.02 &    36.1 & -2.559 & 1 & 3 &  A0pEuCrSi(Sr)\\ 
 150125205  & 29646      & ROT    &   9594 &   1.76 &  819.61 &    35.7 & -2.671 & 1 & 3 &  A1IV         \\ 
 150250959  & 44532      & ROT    &   8072 &   1.21 &  451.68 &    35.8 & -2.073 & 1 &   &  A2V          \\
 160644410  & 131461     & ROT    &   8396 &   1.52 &  615.16 &    36.4 & -1.705 & 1 & 3 &  A0/1V        \\ 
 177284702  & 51581      &        &   7266 &   0.56 &  661.79 &    35.5 & -1.722 & 1 &   &  A8V          \\
 199752613  & 35885      & ROT    &   9566 &   1.09 &  472.60 &    37.0 & -1.332 & 2 &   &  A0           \\
 215256883  & 17864      & ROT    &   9542 &   1.48 &  400.63 &    36.0 & -2.107 & 2 & 3 &  B9.5V        \\ 
 220399820  & 29578      & ACV    &   7415 &   1.40 &  394.36 &    35.7 & -2.312 & 1 &   &  A4SrEuCr     \\
 233164000  & 108346     & ROT    &   9522 &   1.42 &  916.76 &    35.6 & -2.548 & 1 &   &  kA1hA9mF2    \\
 236003103  & 195984     & ROT    &   9900 &   1.50 &  821.51 &    36.0 & -2.317 & 2 &   &  A0V          \\
 248430494  & 33190      & ROT    &  15100 &   2.17 &  447.91 &    36.5 & -2.499 & 1 &   &  B8V          \\
 248992635  & 33819      &        &   8735 &   1.26 &  453.52 &    35.8 & -2.164 & 1 &   &  A0V          \\
 252834311  & 20842      & ROT    &   9900 &   1.39 &  804.06 &    35.4 & -2.643 & 1 &   &  A0Va+        \\
 256749693  & 191174     & ROT    &   9170 &   1.40 &  699.39 &    35.7 & -2.325 & 1 & 2 &  A2II-III     \\ 
 260416268  & 45229      &        &   7537 &   1.29 &  336.89 &    35.8 & -2.291 & 2 & 2 &  kA2hA7VmA7   \\ 
 264593064  & 35134      & ROT    &   8193 &   1.49 &  487.26 &    35.6 & -2.658 & 6 & 3 &  A2V          \\ 
 264683456  & 36030      & ROT    &   8992 &   1.70 &  488.89 &    36.6 & -2.123 & 1 & 6 &  A0           \\ 
 269833435  & 196816     &        &   8129 &   1.02 &  329.45 &    35.0 & -2.722 & 1 &   &  A3/5III      \\
 280965566  & 83719      & ROT    &   7992 &   1.81 &  559.25 &    37.0 & -1.866 & 1 & 3 &  A0V          \\ 
 284084463  & 22961      & ACV    &   9650 &   1.37 &  810.20 &    36.2 & -2.100 & 1 &   &  A1pSr        \\
 287178418  & 86001      & ROT    &   7749 &   1.13 &  856.61 &    35.9 & -2.049 & 1 &   &  A2V          \\
 287329624  & 57642      & ROT    &   6900 &   0.84 &  478.19 &    35.2 & -2.380 & 1 &   &  A8IV/V       \\
 299899924  & 54682      & ROT    &   7404 &   1.50 &  488.61 &    36.4 & -2.044 & 4 &   &  A0V          \\
 301749125  & 155056     & ROT    &   9241 &   1.39 &  664.92 &    35.8 & -2.247 & 2 &   &  A2V          \\
 313942295  & 170868     & ROT    &  10809 &   2.88 &  656.15 &    38.5 & -1.601 & 1 & 7 &  B8/A1        \\ 
 324207960  & 169484     & ROT    &   7175 &   1.84 &  656.45 &    36.9 & -1.558 & 1 & 3 &  A8/9III/IV   \\ 
 324892747  & 173842     &        &   7555 &   1.33 &  664.99 &    36.8 & -1.974 & 1 &   &  A7IV         \\
 327136878  & 9622       & ROT    &   5978 &   0.64 &  814.03 &    34.8 & -2.341 & 1 &   &  A0pSi?       \\
 327724630  & 209468     & ROT    &   8906 &   1.48 &  341.54 &    35.5 & -2.619 & 1 &   &  A1V          \\
 332659885  & 26624      & ROT    &   7960 &   1.14 &  452.09 &    35.2 & -2.447 & 2 &   &  A2/3V        \\
 337220792  & 20769      & ROT    &   9631 &   1.35 &  423.21 &    36.0 & -2.161 & 1 &   &  A0V          \\
 349193923  & 56911      & ROT    &   9624 &   1.47 &  521.06 &    36.0 & -2.310 & 2 &   &  A0Vs         \\
 352939640  & 25553      & ROT    &   9900 &   1.43 &  827.25 &    35.7 & -2.430 & 2 &   &  A0V          \\ 
 357633579  & 79490      & ROT    &   8063 &   1.18 &  552.43 &    36.3 & -2.207 & 2 & 3 &  A1/2V        \\ 
 358467237  & NGC2516 4  & ROT    &   6379 &   0.95 &  333.98 &    36.3 & -1.815 & 2 & 8 &  A7III        \\ 
 360020620  & 190833     & ROT    &   9900 &   1.28 &  700.05 &    37.0 & -1.261 &10 & 3 &  A0V          \\ 
 393389739  & 43881      & ROT    &   9123 &   1.32 &  493.71 &    36.0 & -2.642 & 1 &   &  A2V          \\
 395007683  & 97049      & ROT    &   9034 &   1.19 &  646.65 &    35.4 & -2.209 & 1 &   &  A2V          \\
 404477098  & 15527      & ROT    &   6929 &   1.22 &  423.25 &    37.0 & -1.343 & 1 &   &  A9V          \\
 407825808  & 163837     & ROT    &   7023 &   0.83 &  627.18 &    35.5 & -2.103 & 1 & 9 &  A9V          \\ 
 409135458  & 56832      & ROT    &   7495 &   1.53 &  492.11 &    36.7 & -2.164 & 1 &   &  A1II/III     \\
 426452677  & 143474     & ROT    &   7620 &   1.29 &  636.71 &    36.0 & -2.109 & 1 & 2 &  A5IVs        \\ 
 427393202  & 294262     & ROT    &   6962 &   1.71 &  478.46 &    36.8 & -2.057 & 1 & 3 &  A0           \\ 
 427458366  & 290674     & ROT    &   8416 &   1.53 &  481.88 &    35.9 & -2.347 & 1 &   &  A0V          \\
 434109154  & GD 1214    &        &   8680 &   1.81 &  369.38 &    38.2 & -0.793 & 1 &   &               \\ 
 438598966  & 116649     & ROT    &   8200 &   1.22 &  608.60 &    35.8 & -2.506 & 1 &   &  A0V          \\
 440863421  & 131885     & ROT    &   8793 &   1.47 &  616.47 &    35.3 & -2.882 & 2 &   &  A0V          \\
 442926107  & 35308      & ROT    &   8631 &   1.42 &  455.54 &    36.2 & -2.431 & 2 &   &  A0V          \\
 443316662  & 45341      & ROT    &   7159 &   1.13 &  474.24 &    35.7 & -2.591 & 2 &   &  A2           \\
 452468734  & 80950      & ROT    &  10069 &   1.61 &  592.16 &    35.2 & -2.776 & 1 &   &  A0V          \\
 459786991  & 82861      & ROT    &   7537 &   1.29 &  870.87 &    35.9 & -2.510 & 5 & 3 &  kA2mF0       \\ 
\hline                     
\multicolumn{11}{l}{References:}\\   
\multicolumn{11}{l}{1- \citet{Lo2014}; 2 - \citet{Schroder2007}; 3 - \citet{Voges1999}; 4 -
\citet{Evans2013}; 5 - \citet{Makarov2003};}\\
\multicolumn{11}{l}{6 - \citet{Evans2020}; 7 - \citet{Berghoefer1996}; 8 - \citet{Marino2006}; 9 -
\citet{Voges2000}}\\
\hline
\end{tabular}
\endgroup
}
\end{center}
\end{table*}

\section{Conclusion}

It is demonstrated that the periodic light variations seen in about
40\,percent of A and B stars is consistent with rotational modulation.  The 
expected relationship between the estimated equatorial rotation velocity, $v$, 
and the projected rotation velocity, $v\sin i$, is confirmed for B stars in 
three effective temperature ranges.  Furthermore, the detailed distribution of 
$v$ matches the distribution of $v \sin i$ even for the hottest B stars. 

\citet{Balona2020b} proposed a model for Be stars in which the energy
released by flares and magnetic reconnection, in conjunction with rapid
rotation, ejects gas which is trapped in two diametrically opposite
locations defined by the intersection of the geographic and magnetic
equators.  This gas eventually dissipates into the circumstellar disc.  This
model appears to explain most of the known characteristics of Be stars. 

In this analysis Be were excluded, even though they appear to show
rotational modulation \citep{Balona2020b}. It turns out that the rotational
light amplitudes are an order of magnitude larger than in non-Be stars. 
Rotational modulation in Be stars is, after all, detectable from the ground
which is not true of most A and B stars.  The large amplitude in Be stars
may possibly be a result of circumstellar clouds rather than starspots
\citep{Balona2020b}.  Due to their rapid rotation, inclusion of Be stars
will distort the true velocity distribution.  A full discussion of activity 
and rotation in Be stars will be presented elsewhere.

The nature of the presumed starspots responsible for rotational modulation is
not known.  The idea of magnetic field generation in subsurface convective 
zones first postulated by \citet{Cantiello2009, Cantiello2011, Cantiello2011b,
Cantiello2019} seems to be quite promising.  However, one would have
expected that starspots should occur only within certain effective temperature 
ranges depending on the ionization species.  

According to \citet{Cantiello2019}, the largest effects are caused by a  
convective layer driven by second helium ionization.  The amplitude of surface 
magnetic fields and their associated photometric variability are expected to 
decrease with increasing stellar mass and surface temperature, so that 
magnetic spots and their observational effects should be much harder to detect 
in late B-type stars.  This is clearly not the case, since the fraction of 
late B stars showing rotational modulation is about the same as in A stars 
(Table\,\ref{tab}). In fact, the fraction stays about the same at 
30--40\,percent for all stars in the upper main sequence.  

Another problem is that sub-surface convection predicts the creation of bright 
spots.  We know that spots on the Sun are dark, and this seems to be true of 
solar-type stars as well.  Since rotational modulation is present for the
full range of main sequence stars, there must be a transition between dark
spots and bright spots around early F or late A.  As a result, one might 
expect a decrease in the numbers of stars with rotational modulation in this 
spectral type range.  This does not seem to be the case unless the transition 
is very sharp.

An alternative mechanism proposed many years ago involves the interaction
between magnetic fields, convective flows and differential rotation.  A dynamo 
cycle operating on differential rotation in stellar radiative interiors was 
described by \citet{Spruit1999, Spruit2002} and \citet{Maeder2004} (see also
\citealt{Braithwaite2017}). In this theory, a magnetic instability in the 
toroidal field wound up by differential rotation replaces the role of 
convection in closing the field amplification loop in conventional dynamo 
theory.  It is possible that completely stable radiative envelopes do not 
exist and that turbulence generated by differential rotation may lead to 
surface magnetic fields capable of forming conventional dark spots.

Examination of {\em TESS} A and B stars has led to the detection of 68 new
early-type flare stars, doubling the total number.  These include some Ap
and Am stars, with some stars being X-ray sources. If starspots are deemed 
to be present, in A and B stars, then there should be no barrier to accepting 
that flares may be generated by magnetic reconnection, as in the Sun and cool 
stars. Indeed, it then becomes necessary to provide reasons why flares should 
{\em not} be generated in A and B stars.

The apparent magnitudes of the A stars are 4--10\,mag with a median 8.2\,mag. 
A cool foreground dwarf of comparable brightness, or even significantly 
fainter, will be one of the nearest stars and well documented.  Flares
originating in a cool foreground star can be excluded.  One can also exclude
foreground F or G giants because the combined colour would not be that of an
A star (in any case the stars all have A or B spectral classifications). 

A cool K or M binary companion can also be excluded because all 110 detected
flares have energies considerably larger than that of the largest flare ever
seen in a cool dwarf.  There are two remaining possibilities: the flare arises 
in magnetic reconnection involving the A star and a close companion, or solely 
on the A star itself.  Either way it means that a significant magnetic field 
must be present on the A star.  We therefore return to the original problem
regarding the presence of magnetic fields in radiative envelopes. 

For further progress it will be important to design observations which might
lead to resolving the problem of whether the spots are bright or dark. 
Further high-resolution spectroscopy of A or B stars, as performed by
\citet{Bohm2015} on Vega would be important to place limits on the size and
distribution of the spots.  It would also be important to obtain
time-series spectroscopy on A/B flare stars to determine possible
interacting companions.

\section*{Acknowledgments}

I wish to thank the National Research Foundation of South Africa for 
financial support. I also thank the {\it TESS} Asteroseismic Science 
Operations Center (TASOC).

Funding for the {\it TESS} mission is provided by the NASA Explorer Program. 
Funding for the {\it TESS} Asteroseismic Science Operations Centre is provided 
by the Danish National Research Foundation (Grant agreement no.: DNRF106), 
ESA PRODEX (PEA 4000119301) and Stellar Astrophysics Centre (SAC) at Aarhus 
University. 

This work has made use of data from the European Space Agency (ESA) mission 
Gaia, processed by the Gaia Data Processing and Analysis Consortium (DPAC).
Funding for the DPAC has been provided by national institutions, in particular 
the institutions participating in the Gaia Multilateral Agreement.  

This research has made use of the SIMBAD database, operated at CDS, 
Strasbourg, France.  Data were obtained from the Mikulski Archive for Space 
Telescopes (MAST).  STScI is operated by the Association of Universities for 
Research in Astronomy, Inc., under NASA contract NAS5-2655.

\bibliographystyle{mnras}
\bibliography{solcon}

\label{lastpage}

\end{document}